# Spreadsheets: Aiming the Accountant's Hammer to Hit the Nail on the Head

Mbwana Alliy–Microsoft and Patty Brown–Two Degrees
One Microsoft Way
Redmond, WA 98052 - USA
malliy@microsoft.com
pattyb@twodegrees.com

## ABSTRACT

*Accounting and Finance (A&F) Professionals are arguably the most loyal and concentrated population of spreadsheet users. The work that they perform in spreadsheets has the most significant impact on financial data and business processes within global organizations today.*

*Spreadsheets offer the flexibility and ease of use of a desktop application, combined with the power to perform complex data analysis. They are also the lowest cost business IT tool when stacked up against other functional tools. As a result, spreadsheets are used to support critical business processes in most organizations. In fact, research indicates that over half of financial management reporting is performed with spreadsheets by an accounting and finance professional.*

*A disparity exists in the business world between the importance of spreadsheets on financial data (created by A&F Professionals) and the resources devoted to:*

- *The development and oversight of global spreadsheet standards*
- *A recognized and accredited certification in spreadsheet proficiency*
- *Corporate sponsored and required training*
- *Awareness of emerging technologies as it relates to spreadsheet use*

*This management paper focuses on the current topics relevant to the largest user group (A&F Professionals) of the most widely used financial software application, spreadsheets, also known as the accountant's hammer.*

## 1. INTRODUCTION

In reading the latest topics discussed on any spreadsheet related internet discussion board, it is apparent what is top of mind for many A&F Professionals who are required to use spreadsheets in their line of work:

- Spreadsheet controls in a compliance environment
- Non-existence of recognized spreadsheet standards
- Inconsistent training and certification
- Lack of awareness of emerging spreadsheet technologies and capabilities

These topics are consistent with the many articles and papers that have been previously published and presented. Productivity is the underlying mission and theme in each of these topics, which continue to be challenges for A&F professionals. Conversely, increased awareness of many of these topics has prompted development in technologies to elevate spreadsheet quality and productivity.

Each topic is discussed throughout this paper. Since much has been written on the details of these topics, the sections below describe only briefly the background of each topic and will instead focus more on updated findings and solutions.

## 2. CONTROLS AND COMPLIANCE – OLD CHALLENGE #1

Organizations today are under considerable regulatory pressure to ensure that financial reporting processes are both transparent and well-documented. Most publicly-traded companies as well as industry and local regulators around the world are impacted by this recent legislation.

Most companies rely on spreadsheets for financial reporting and operational processes. As the complexity and importance of a spreadsheet increases, so too does the cost of errors and innaproriate disclosures of data. The above legislation increases the focus on controls related to the development and maintenance of spreadsheets.

### 2.1 Controls and Compliance - New Solutions to this Old Challenge

Risks to spreadsheet compliance can be mitigated by implementing controls on important elements of business-critical spreadsheets to allow only authorized users to view content, make changes, and share information. Tools and technology are available to organization to take full advantage of to help fulfill compliance requirements.

In response to regulation, capabilities have been developed in the last year in widely used spreadsheet programs to be used in conjunction with a sound compliance strategy to address compliance challenges with spreadsheet use. These capabilities include:

**Preventing Unauthorized Access to Spreadsheets**

Spreadsheets offers technolgies for helping to secure critical spreadsheets from unauthorized access and modification on both the client and server related to work book encryption, permission, sharing, better control over who can open, copy, print, or forward information

**Managing and Monitoring Spreadsheet Changes**

A sound compliance strategy will include some level of on-going change management and monitoring for critical spreadsheets. Content management tools can facilitate this process through the versioning, auditing, and workflow capabilities which allow users to better manage important spreadsheets and documents without sacrificing productivity.

**Retaining and Archiving Spreadsheets**

The ability to archive spreadsheets is just one component of a larger records management process that includes the collection, management, and disposal of corporate records in a consistent and uniform manner based on the company's policies. Advances in new record management capabilities within content management tools help companies ensure that vital corporate records, including critical spreadsheets, are properly retained for legal, compliance, and business purposes and then properly disposed of when no longer needed.

**Developing Robust Spreadsheet Models**

Newer spreadsheet versions can be used to create a robust spreadsheet model that meets compliance challenges and enhances productivity. For example, the following capabilities in Excel 2007 can help an organisation deploy spreadsheet models that make it easier to become, and stay, compliant.

- Cell Styles
- Lock important cells
- Using Excel Tables to reduce errors
- Defined Names
- Formula auditing tools

**2.2 Controls and Compliance - New Findings to this Old Challenge**

Despite key changes in spreadsheet applications designed to alleviate risk, user inertia is prevalent and a challenge to overcome.

Before these new technologies, users invested their time to create work arounds in spreadsheets to ensure compliance with regulations. They became good at these work arounds and stuck to it. It may not be the best way to manage compliance efforts but users can work *faster*, even if it is a less efficient application. They have become an expert. The result is users are not using these new technologies that were developed to improve productivity and reduce control risk.

This mindset of user inertia can be alleviated with increased attention to the spreadsheet development profession in areas such as spreadsheet standards, training, certification and awareness of new technologies, which are all further discussed below. Increased attention would result in users being better informed and more productive and the quality of spreadsheet output is improved.

**3. GLOBAL SPREADSHEET STANDARDS – OLD CHALLENGE #2**

Even in a highly regulated society, regulations around spreadsheet use has not made it's way through US legislation or in other countries for that matter, despite the increasing reliance on spreadsheets in financial reporting. As a result, many users create their own spreadsheet standards, use standards created by other non authoritative sources or worst case, don't use any standards at all. The consequence is in inconsistent use of spreadsheet use which poses a risk to spreadsheet integrity.

**3.1 Standards - New Findings to this Old Challenge**

Standardization is a commonly accepted approach to lowering costs and realizing process efficiencies. Examples of standards related to other expertise include Generally Accepted Auditing Standards (GAAS) for auditors of public accounting firms; Financial and Reporting Standards issued by the Financial Accounting Standards Board and the American Bar Association Criminal Justice Standards for criminal justice practitioners and policymakers.

Other examples of standards and oversight include the Public Company Accounting Oversight Board for SOX, the XBRL International Consortium for eXtensibile Business

Reporting Language (XBRL), the ERB Consortium for Enhanced Business Reporting (EBR) and RosettaNet for global supply chain standards.

BPM Global, a company based in Australia, has developed their own spreadsheet standards and created their own Spreadsheet Standards Review Board to provide oversight. Although not universally recognized, these comprehensive spreadsheet standards are unique as compared to what exists today and provide a resourceful framework and common language for their own spreadsheet community that follow it.

### 3.2 Standards - New Solutions to this Old Challenge

Other standards exist in areas that have major impact on businesses. So, why not global spreadsheet standards and oversight from a global standards board? Opponents believe standards would limit creativity. Proponents of spreadsheet standards believe they would increase awareness of spreadsheet use, promote spreadsheet integrity and productivity and positively impact the bottom line by reducing costs.

Global spreadsheet standards have the potential to increase confidence in financial reporting, strengthen the integrity of spreadsheets, give users a standard baseline to strive for and encourage consistent use of spreadsheets.

One area that gets spreadsheets closer to standardization is XBRL technology, which is further discussed below.

### 4. END USER TRAINING – OLD CHALLENGE #3

Excel has radically evolved from the first release in 1985 to the newest version released in 2007. It has also advanced to be a primary tool that organizations rely on for their financial reporting. Yet, many Excel users have not voluntarily taken formal training to stay current on newer versions nor are they required by their employers to take any initial or continuing education as it relates to Excel. The reason for this is not because there is a short supply of training available. Typing in "Excel training" in any internet search engine will return tens of thousands of results. It can be overwhelming to spreadsheet users to determine what is effective and relevant. User inertia is hard to overcome and contributes to low participation in training.

The deficiency in end user training in Excel contributes to the inherent risk associated with spreadsheet error.

### 4.1 Training - New Findings to this Old Challenge

Two Degrees, an A&F professional services firm in Seattle, provides formal Excel training to its Principal Associates who provide A&F services to clients. Two Degrees conducted their own study amongst its Principal Associates to determine what type of training in Excel they had received from previous employers.

- Only 5% were *required* by their previous employer to take an Excel training course.
- 25% were *encouraged* and reimbursed by their employer to take an Excel training course.
- 100% were self taught and of those, 20% also took an Excel training course.

If these percentages are representative of the spreadsheet user population, it illustrates the lack of attention to end user training of the most used application relied upon for financial reporting.

**4.2 Training - New Solutions to this Old Challenge**

Along the same lines as global spreadsheet standards is the idea of a global accredited spreadsheet training program where topics are aligned with what is important and relevant to the spreadsheet user community.  Required continuing education would also benefit any training program so that users are updated on major changes in spreadsheet use.  To take training to the next level, certification and expertise assessment gives users a level to strive for and is further discussed below.

**5.   EXPERTISE ASSESSMENT AND CERTIFICATION – OLD CHALLENGE #4**

Much has been written about the human tendency to be overconfident in one's ability.  Experiments and surveys have repeatedly shown that most people believe that they possess attributes that are better or more desirable than average.  This is no different amongst spreadsheet users, regardless of the impact spreadsheets have on financial reporting.  There is not a recognized certification to ensure that spreadsheet users are aware or educated on very beneficial features in spreadsheets that are now available to reduce spreadsheet risk.  On the other hand, there are many certifications available for other technical skills to achieve quality of performance in using those skills.  A recognized certification on spreadsheet skills would define expertise assessment levels to facilitate identifying, hiring, promoting, and retaining qualified individuals as well as encourages user to strive for higher standards and advance in their spreadsheet skills.  Organizations can benefit from a streamlined recruitment process, greater individual productivity, and increased employee satisfaction.  This will inevitably result in a reduction of spreadsheet risk.

**5.1  Certification – New Findings to this Old Challenge**

Two Degrees surveyed their Principal Associates who are considered super users of spreadsheets, and who are described by clients as top performers, to define their level of Excel expertise as Beginner, Intermediate or Advanced.  No definition of these levels was given so that each person was left to come to their own conclusion on how each level was defined.  In each case, all users described their skills as Intermediate or Advanced.  When each of the Principal Associates were asked to take the Microsoft Certified Application Specialist (MCAS) exam, which tests on broad features in Excel 2007, only 20% passed the exam on the first try.  Those that did not pass the first time were asked where they fell short and the most common answers were that they were not familiar with all of the features that Excel 2007 offered or how the user-interface had evolved.  They based their expertise level on older versions of Excel and the specific features they use in Excel as it relates to accounting and finance.

Excel users generally classify their level of expertise based on older versions when in fact they are not proficient in the more sophisticated features offered in newer versions that have addressed past issues such as controls and compliance.

**5.2  Certification – New Solutions to this Old Challenge**

Two Degrees created the Excel Black Belt certification for its internal employees to close the gap in expertise assessment.   Although the certification is not accredited, it does provide definition, consistency and expectation internally in order to assess technical proficiency and expertise by evaluating comprehension of Excel specifically of an A&F professional. With the certification their employees are able to provide exceptional client service in older and newer versions of Excel, including spreadsheet compliance, design, use, collaboration, sharing and training using a holistic view as illustrated below:

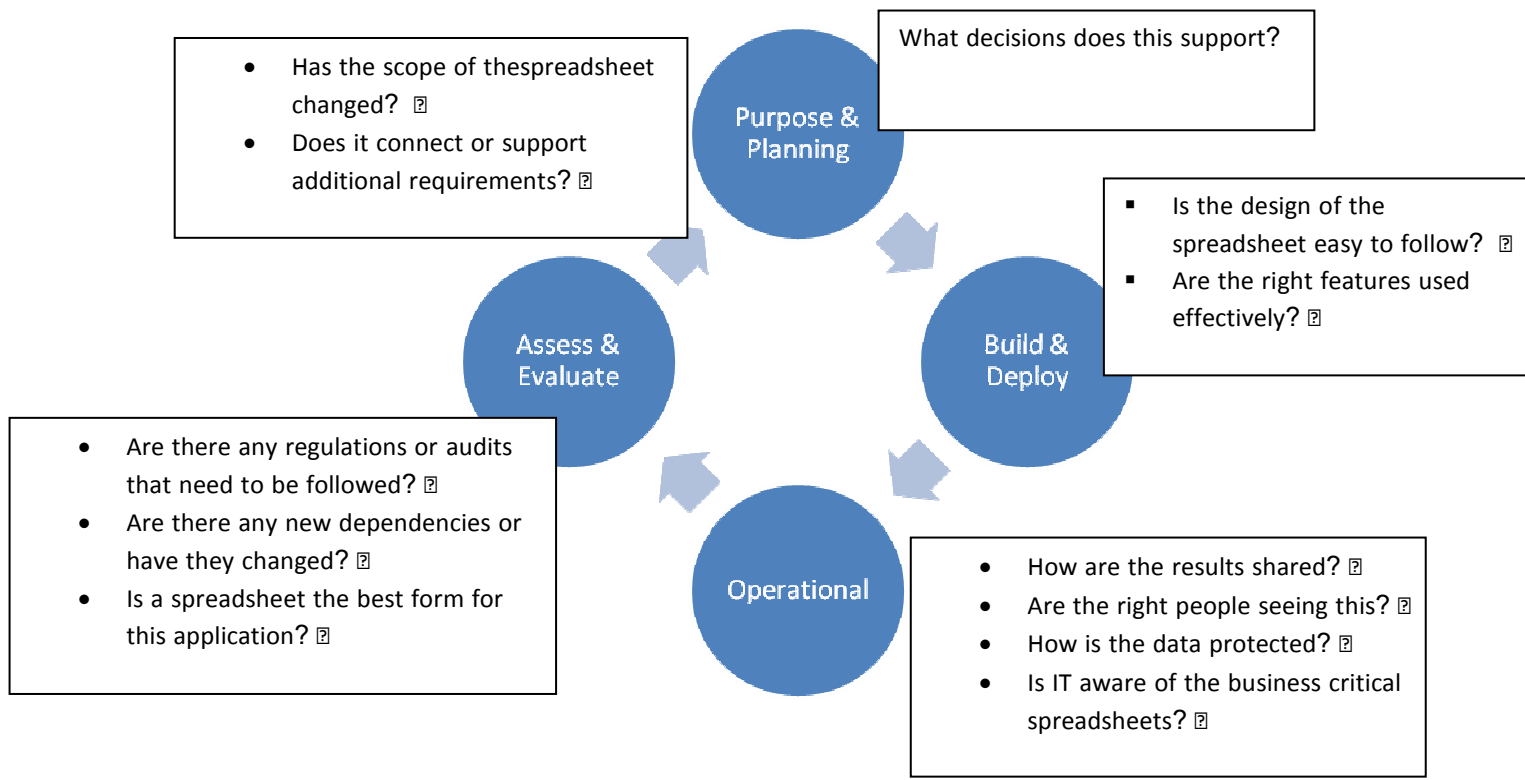

The Excel Black Belt certification provides professionals the opportunity to be advanced users of Excel, learn best practices and teach and educate users to ensure excellence in spreadsheet use.

## 6. AWARENESS OF EMERGING TECHNOLOGIES – OLD CHALLENGE #5

Recent advances in technology address pervasive spreadsheet issues and alleviate risks associated with spreadsheets. Lack of awareness of what is available is an on-going frustration for users and ties in to lack of training. Following are some of the most known examples of advancements.

### 6.1 XBRL

XBRL technology has received considerable attention by the Securities and Exchange Commission (SEC) because of the benefits it has on financial reporting. XBRL is a language for the electronic communication of financial data which is revolutionising financial reporting around the world. It provides benefits in the preparation, analysis and communication of financial information. It offers cost savings, greater efficiency and improved accuracy and reliability to those supplying or using financial data. What XBRL does for financial reporting is often compared to what bar coding does for retailers: tracks, simplifies, standardizes and manages data and processes. To highlight the importance XBRL has on financial reporting, the SEC announced in May 2008 a proposal that would require public companies in the US to use interactive data to file financial statements beginning next year.

As it relates to spreadsheets, XBRL is expected to alleviate reliance on human input and manipulation thus reducing the risk of human error in financial reporting. It can also improve spreadsheet controls by standardizing the information model contained within the spreadsheet and automating the currently manual steps. XBRL-enabled spreadsheets function like virtual dashboards, allowing consumers to access, use and report on any information accessible in a standardized manner, such as XBRL's Global Ledger Framework and other XBRL taxonomies vs. using just a spreadsheet. Using XBRL in spreadsheets will only boost the power of spreadsheets and at the same time enhance controls.

### 6.2 Excel

Excel 2007, Excel Services in conjunction with SharePoint Server 2007 and other technologies such as Information Rights Management address many issues from a controls and compliance standpoint.

### 6.3 Spreadsheet Add-Ins

A spreadsheet add-in is a file that spreadsheet applications load at start up. They add additional functionality and increase the power to spreadsheets. Spreadsheet applications include a variety of add-ins and many third-party add-ins are available. Internet discussion boards are loaded with "how to" questions and where to find information. Some of the answers lie in an add-in. Increased awareness of what add-ins is available would hugely benefit spreadsheet users in meeting their needs and improving spreadsheet quality. This is another example of where universal training or a standards board could benefit the spreadsheet user community to heighten awareness on what is available to them.

### 6.4 Embracing Technology

User inertia is a common theme throughout the paper as it relates to embracing technology. As is the case of XBRL, a regulatory agency or standards board can intercept as a means to compel users to embrace new technologies that are meant to improve the quality of spreadsheet output that is so largely relied upon and at the very least, create awareness so spreadsheet users are aware of what is available to them.

### 7. CONCLUSION

The risks associated with spreadsheet use are an old and increasingly discussed topic. This paper has laid out persistent challenges and as well as advances made in reducing spreadsheet risk. Organizations such as EUSPRIG and Spreadsheet Safe in Europe are already contributing to higher standards in spreadsheet use. By globally uniting the spreadsheet user community in the areas of standards, productivity, training, certification and awareness, the former challenges of spreadsheet risk can be significantly reduced and the pursuit of spreadsheet excellence achieved.